\documentclass[prd,11pt,preprintnumbers,nofootinbib,floatfix]{revtex4-1}
\usepackage{mathptmx}
\usepackage[english]{babel}
\usepackage{amssymb}
\usepackage{amsfonts}
\usepackage{amsmath}
\usepackage{eucal}
\usepackage{graphicx,color}
\usepackage{epsfig}
\newcommand{\be}{\begin{equation}} \newcommand{\ee}{\end{equation}}
\newcommand{\bea}{\begin{eqnarray}} \newcommand{\eea}{\end{eqnarray}}
\newcommand{\beann}{\begin{eqnarray*}}  \newcommand{\eeann}{\end{eqnarray*}}
\newcommand{\bfig}{\begin{figure}} \newcommand{\efig}{\end{figure}}
\newcommand{\ba}{\begin{array}} \newcommand{\ea}{\end{array}}
\newcommand{\bcen}{\begin{center}} \newcommand{\ecen}{\end{center}}
\newcommand{\btab}{\begin{tabular}} \newcommand{\etab}{\end{tabular}}

\renewcommand{\Re}{\mathop{\rm Re}}   
\newtheorem{Proposition}{Proposition}[section]

\newtheorem{Theorem}{Theorem}[section]
\newtheorem{Lemma}{Lemma}[section]
\newtheorem{Corrolary}{Corrolary}[section]

\newcommand{\bp}{\begin{Proposition}}	\newcommand{\ep}{\end{Proposition}}
\newcommand{\bt}{\begin{Theorem}}	\newcommand{\et}{\end{Theorem}}
\newcommand{\bl}{\begin{Lemma}}		\newcommand{\el}{\end{Lemma}}
\newcommand{\bc}{\begin{Corrolary}}	\newcommand{\ec}{\end{Corrolary}}
\begin{document}

\title{Anomaly related transport of Weyl fermions for Weyl semi-metals}

\author{Karl Landsteiner}
\affiliation{Instituto de F\'{\i}sica Te\'orica UAM/CSIC, C/ Nicol\'as Cabrera
13-15,\\
Universidad Aut\'onoma de Madrid, Cantoblanco, 28049 Madrid, Spain}

\begin{abstract}

We present a field theoretical model of anomalous transport in Weyl
semi-metals. 
We calculate the Chiral Magnetic and Chiral Vortical Effect in the electric,
axial (valley) and energy current. 
Our findings coincide with the results of a recent analysis using kinetic theory
in the bulk of the material.
We point out that the kinetic currents have to be identified with the {\em
covariant} currents in quantum field
theory. These currents are anomalous and the CME appears as anomalous charge
creation/annihilation at the edges
of the Weyl semi-metal.
We discuss a possible simultaneous experimental test of the chiral magnetic and
the chiral vortical effect sensitive
to the temperature dependence induce by the gravitational contribution to the
axial anomaly.
 \end{abstract}

\preprint{IFT-UAM/CSIC-13-074}
\preprint{INT-PUB-13-022} 
\maketitle
%
\section{Introduction.}
\label{sec:intro}

Weyl semi-metals are materials whose electronic structure features
quasi-particles 
that are locally in momentum space well described by massless chiral fermions
via
the Weyl equation. Realizations of such materials have been
suggested in \cite{wanturneretal, YangYuetal, BurkovBalents}.
One of the most characteristic  properties of the quantum theory
of chiral fermions is the presence of chiral
anomalies \cite{Adler:1969,Bell:1969,Delbourgo:1972xb,Eguchi:1976db} 
(see \cite{Bertlmann:1996xk,Fujikawa:2004cx} for extensive reviews). Not all
symmetries that are
present in the classical Hamiltonian or Lagrangian are really present on the
quantum
level. How the anomaly can be realized effectively in crystal lattices has been 
explained some time ago in \cite{Nielsen:1983rb}. From a modern point of view
the essential
feature is that locally around the tip of the Weyl cone in momentum space the
electron
wave function is subject to a non-trivial Berry phase \cite{Volovik:2003fe,
Son:2012wh}
with a Berry curvature of integer flux $k$ around the cone. The anomaly in each
Weyl cone takes the form
\begin{equation}\label{eq:anomaly}
 \partial_\mu J^\mu = \frac{k}{4\pi^2} \vec{E}.\Vec{B}\,.
\end{equation}
A physically insightful derivation of this equation has been presented in \cite{Nielsen:1983rb} 
in the context of Weyl fermions in a crystal. It can be phrase in the following
way: in a magnetic field the spectrum of a Weyl fermion splits in to Landau levels
with the lowest level behaving as a chiral fermion in one dimension. Its momentum
has to be aligned (or anti-aligned) with the magnetic field. If we switch on
also an electric field parallel to the magnetic one the Lorentz force acting
on the fermions in the lowest Landau level implies $\dot p = \pm E$ depending
on chirality, where $p$ is the momentum
and $E$ the electric field. This implies that the Fermi momentum $p_F$ is shifted. The
change in the density of states at the Fermi level is given by $dp_F/(2\pi)$ multiplied with the
degeneracy of the lowest Landau level, winch is $B/(2\pi)$ Therefore the number of
states changes with time as $\dot n = \pm EB/(4\pi^2)$, which is nothing but a
non-covariant version of the anomaly equation (\ref{eq:anomaly}). Note that this
derivation of the anomaly works separately for each Weyl cone.
The Nielsen-Ninomiya no-go theorem implies furthermore that the total sum over the Berry
fluxes in 
the Brillouin zone vanishes. So the simplest realization is a model with 
$k\in\{+,-\}$, i.e. one right- and one left-handed chiral fermion.
It also implies that in metals there is no 'real'
anomaly 
but the (axial) anomaly is simulated by electrons moving from one Weyl cone to
another one
of opposite Berry flux \cite{Nielsen:1983rb}.  Equation
(\ref{eq:anomaly}) describes the rate of this effective chirality change
in external electric and magnetic fields. The description as Weyl fermions
is valid only inside a limited energy range. An electron present in one Weyl
cone can scatter outside this energy range and reappear in the other Weyl
cone of opposite chirality. For the effective description in terms of Weyl
fermions this looks as if one fermion had changed its chirality. Such a 
chirality changing process
is possible even in the absence of external electric and magnetic fields and
represents a tree-level breaking of the axial symmetry akin to a mass term
for a Dirac fermion.

At finite temperature and density anomalies are intimately related to the
existence of
non-dissipative transport phenomena. More precisely a magnetic field induces a
current
via the chiral magnetic effect (CME) and a vortex or rotation of the fluid or
gas of chiral
fermions also induces a current via the chiral vortical effect (CVE)
\cite{Vilenkin:1979ui, Vilenkin:1980fu, Vilenkin:1980zv,Giovannini:1997eg,
Alekseev:1998ds, 
Son:2004tq, Newman:2005as, Newman:2005hd, Kharzeev:2007jp, Kharzeev:2007tn,
Erdmenger:2008rm, Banerjee:2008th, 
Fukushima:2008xe, Kharzeev:2009pj, Son:2009tf, Landsteiner:2011cp, Nair:2011mk,
Loganayagam:2012pz} 
(see also the recent reviews \cite{Zhitnitsky:2012ej, Fukushima:2012vr,
Basar:2012gm, Zakharov:2012vv, Gorsky:2013pwa, 
Hoyos:2013qwa, Buividovich:2012kq, Yamamoto:2012bi, Landsteiner:2012kd} )
Generalizations to arbitrary dimensions have been discussed in
\cite{Loganayagam:2011mu}. The hydrodynamic approach has been generalized in
\cite{Neiman:2010zi, Sadofyev:2010pr}. Since the effects are non-dissipative
they can
also be obtained from effective action approaches \cite{Banerjee:2012iz,
Jensen:2012kj}
or using Ward identities \cite{Jensen:2012jy}. 
In the context of high energy physics these effects are nowadays well
established.
They have been found also via lattice QCD
\cite{Buividovich:2009wi, Buividovich:2010tn, Yamamoto:2011ks, Braguta:2013loa,
Buividovich:2013jba} 
and via holographic methods \cite{Newman:2005hd, Erdmenger:2008rm,
Banerjee:2008th, Yee:2009vw, Rebhan:2009vc,
Gynther:2010ed, Amado:2011zx, Hoyos:2013qwa, Kalaydzhyan:2011vx,
Landsteiner:2011iq}.

It is however still an open question if they really lead to observable effects
in
heavy ion
collisions, such as charge separation \cite{Kharzeev:2007jp}, a chiral magnetic
wave \cite{Kharzeev:2010gd, Burnier:2011bf} or enhanced production of high spin
baryons
\cite{KerenZur:2010zw}.
A recent review of the experimental situation at heavy ion collisions can be
found in \cite{Bzdak:2012ia}.

In the context of Weyl semi-metals the theoretical situation seems less clear. 
Statements of existence of the CME in Weyl semi-metals are contrasted with some 
explicit calculations that
see no such effect \cite{Zyuzin:2012tv, Zyuzin:2012vn, ChenWuBurkov, goswami, 
Liu, adolfo, Zhou, franz, Basar:2013iaa, Goswami:2013bja}. 
Also different approaches have been put forward to describe anomaly induced
transport
in Weyl semi-metals. One is the point of view of effective field theory of
chiral
fermions valid at long wave-length and for excitations around the Fermi surface.
This can be contrasted with a more down-to earth picture in which there are
simply electrons 
filling a particular band structure. 
The difference between these two points of view is that in the relativistic 
Weyl fermion picture we need to deal with the intricacies of relativistic field
theory, particles
and anti-particles and a as a result a complicated vacuum. In contrast,
electrons are
on a fundamental level simply electrons without any anti-particles even if they
partially fill some non-standard band structures.

The purpose of this article is to develop a field theoretical model. 
In particular we want to connect the well known anomaly related transport 
formulas for relativistic Weyl fermions with the results of  kinetic theory of 
electrons at Weyl nodes
\cite{Son:2012wh, Basar:2013iaa, Stephanov:2012ki, Zahed:2012yu, Chen:2012ca,
Gao:2012ix, Basar:2013qia}. 
Using a field theoretical model forces us however also
to go substantially beyond the level of discussion available in kinetic theory.
One of
the central issues is what precisely does the formula (\ref{eq:anomaly})
correspond
to in field theoretical language? Anomalies come in two formulations, one as
{\em covariant} anomaly and the other as {\em consistent} anomaly
\cite{Bardeen:1984pm}. Accordingly
one can define covariant and consistent currents. Whereas the divergence of
covariant currents is uniquely defined the divergence of consistent currents
depends on specific regularization schemes. It is this freedom of
regularization,
more precisely, the freedom to modify effective actions by adding local
counterterms,
that allows the definition of an exactly conserved electric current. 
We argue that both formulations of the anomaly, the covariant one and
the consistent one, lead to the same physical outcome: charge separation
at the edges perpendicular to the magnetic field even when there is no
bulk current according to kinetic theory.

We also extend the results to the CME and CVE in the energy current and
prove agreement of our model and kinetic theory. One of the interesting
results is that the temperature dependence related to the presence of 
mixed axial-gravitation anomaly \cite{Delbourgo:1972xb, Eguchi:1976db} 
enters the vortical effect in the energy current. This is the only vector-like 
current, i.e. sum of left-handed and right-handed currents, that is sensitive to
it. We suggest an experimental setup that allows to test the temperature
dependence and
therefore the gravitational contribution to the axial anomaly.

This paper is organized as follows. 
In section \ref{sec:transport} we adapt the well-known formulas for the
anomalous
transport coefficients to the situation present in Weyl semi-metals. 
We then compute the chiral magnetic and chiral vortical conductivities and find
that
our formulas for the covariant current coincide with \cite{Basar:2013iaa}. We
also
compute the chiral conductivities in the energy current leading to the
interesting result that the vortical effect in the energy current is sensitive
to the $T^2$ temperature dependence induced by the gravitational anomaly.

In section \ref{sec:covcons} we emphasize that the currents of the previous
section
have to be understood as the {\em covariant} currents in a quantum field
theoretical
definition. These currents are anomalous and we show that the CME arises as
anomalous charge creation and annihilation at the edges of the Weyl semi-metal.

In section \ref{sec:experiments} we suggest a simple experiment that uses both,
the chiral
vortical and the chiral magnetic effect. Not only does it test the presence of
chiral transport
in general but it also is sensitive to the $T^2$ term related to the
gravitational anomaly.

We present some additional discussion of the results and outlook to future work
in section \ref{sec:discussion}.

\section{Anomalous transport}
\label{sec:transport}

The chiral magnetic effect is only the most prominent example of a larger set of
anomaly related transport coefficients.
The full set of magnetic and vortical effects can be written down as
\cite{Neiman:2010zi}, \cite{Landsteiner:2012kd}, \footnote{These are
one-loop expressions which are known to receive higher loop corrections when
dynamical gauge fields are important
\cite{Hou:2012xg}, \cite{Golkar:2012kb},  \cite{Gorbar:2013upa}. In the case of
Weyl semi-metals these
corrections can be assumed to be small and will be ignored.}
\begin{align}
\label{eq:cmc}
\sigma^B_{ab} &= \frac{1}{4\pi^2} d_{abc} \mu_c \,,\\
\label{eq:cvc}
\sigma^V_{a} &= \sigma^B_{\varepsilon,a} = \frac{1}{8\pi^2} d_{abc}\mu_b\mu_c +
\frac{1}{24} b_a T^2\,,\\
\label{eq:cvce}
\sigma^V_\varepsilon & = \frac{1}{12\pi^2} d_{abc} \mu_a \mu_b \mu_c +
\frac{1}{12} b_a \mu_a T^2\,.
\end{align}
The constants $d_{abc}$ and $b_a$ indicate the presence of gauge and
gravitational
anomalies 
\begin{equation}
D_\mu J^\mu_a = \frac{d_{abc}}{32\pi^2} \epsilon^{\mu\nu\rho\lambda}
F^b_{\mu\nu} F^c_{\rho\lambda} +
\frac{b_a}{768\pi^2} \epsilon^{\mu\nu\rho\lambda}
R^\alpha\,_{\beta\mu\nu}R^\beta\,_{\alpha\rho\lambda} \,.
\end{equation}
These are general expressions valid for a family of symmetries (possibly
non-abelian) labeled by the indices $a,b,c$.
The constants $d_{abc}$ and $b_a$ are 
\begin{align}
d_{abc} &= \text{sTr}\left(T_a T_b T_c  \right)_R -  \text{sTr}\left(T_a T_b
T_c\right)_L\,,\\
b_a &=\text{Tr}  \left( T_a \right)_R -  \text{Tr}\left(T_a\right)_L\,,
\end{align} $\text{sTr}$ stands for a symmetrized trace, $T_a$ is the matrix
generator of the symmetry
labeled by $a$ and the subscripts $R,L$ denote right- and left-handed fermions
respectively.

We would like to emphasize that the gravitational anomaly is not necessarily
related to the presence of space-time curvature.
It rather reflects the fact that in the quantum field theory of chiral fermions
there is an obstruction to define at the same time conserved chiral currents and 
a conserved energy momentum tensor.
This obstruction is manifest 
on the level of triangle diagrams with one chiral current and two energy
momentum tensors even for quantum field theory in flat Minkowski space.

Note in particular that the
temperature dependence in the anomalous conductivities enters via the mixed
gauge-gravitational anomaly coefficient $b_a$. It is worth mentioning 
that the contribution of the gravitational anomaly to transport at first order
in derivatives, such as the responses to magnetic field and vorticity, is surprising
because it is is of higher (fourth) order in derivatives.
How these responses can be obtained in a model independent way using a combination of hydrodynamic and
geometric reasoning was recently shown in \cite{Jensen:2012kj}.

The conductivities defined in equations (\ref{eq:cmc})-(\ref{eq:cvce}) determine
the response of the system in the currents $\vec{J}_a$ and the 
energy current $J^i_\varepsilon = T^{0i}$ in
the presence of a magnetic field $\vec{B}_a$ and rotation measured by the
vorticity $\vec\omega = \frac 1 2 \nabla \times \vec{v}$ with
$\vec{v}$ the vector field describing the rotation. The response is 
summarized in\footnote{The factor of two in the vortical effects arises since
we use the transport coefficients normalized to the gravito-magnetic field in
which $2\omega_i = \epsilon_{ijk} \partial_j h_{0k}$ for the metric fluctuation
$h_{0i}$ coupling to the energy current.}
\begin{align}\label{eq:transport}
 \vec{J}_a = \sigma^B_{ab} \vec{B}_{b} + 2\sigma^V_a \vec{\omega} \,,\\
\label{eq:transporte}
 \vec{J}_\varepsilon = \sigma^B_{\varepsilon,b} \vec{B}_b +
2\sigma^V_\varepsilon \vec{\omega}\,. 
\end{align}


One way of deriving these anomalous conductivities is via the formalism of Kubo
formulae 
\cite{Kharzeev:2009pj, Landsteiner:2012kd}
\begin{align}
  \sigma^B &= \lim\limits_{k_j \rightarrow 0} \sum_{i,k} \epsilon_{ijk} \frac{i}{2k_j} \langle J^i J^k \rangle|_{\omega=0} \,,\\
  \sigma^V &= \lim\limits_{k_j \rightarrow 0}\sum_{i,k} \epsilon_{ijk} \frac{i}{2k_j} \langle J^i J_\epsilon^k \rangle|_{\omega=0} \,,\\
  \sigma^V_\epsilon &= \lim\limits_{k_j \rightarrow 0}\sum_{i,k} \epsilon_{ijk} \frac{i}{2k_j} \langle J_\epsilon^i J_\epsilon^k \rangle|_{\omega=0}
\end{align}

There the expressions arise from the
purely 
finite temperature and finite density parts of the amplitudes (i.e. neglecting
the vacuum
contribution) of current-current and current-energy-momentum tensor two point
functions. 
Without going into details one finds for one chiral (right-handed) fermion
\begin{align}
 \sigma^B &= \frac{1}{4\pi^2} \int_0^\infty dk\;
\left[n_F\left(\frac{k-\mu}{T}\right) - n_F\left(\frac{k+\mu}{T}\right) \right]
= \frac{\mu}{4\pi^2}\,,\\
 \sigma^V = \sigma_\epsilon^B &= \frac{1}{4\pi^2} \int_0^\infty dk\; k
\left[n_F\left(\frac{k-\mu}{T}\right) + n_F\left(\frac{k+\mu}{T}\right) \right]
= \frac{\mu^2}{8\pi^2}+ \frac{T^2}{24}\,,\\
 \sigma^B &= \frac{1}{4\pi^2} \int_0^\infty dk\; k^2
\left[n_F\left(\frac{k-\mu}{T}\right) - n_F\left(\frac{k+\mu}{T}\right) \right]
 = \frac{\mu^3}{12\pi^2}+ \frac{\mu T^2}{12}\,
\end{align}

It is interesting to re-write these formulas in the following way. Let us assume
that $\mu>0$. Then
the argument of the Fermi-Dirac distribution in the first terms in the integrals
(\ref{eq:cmc}-\ref{eq:cvce}) are
negative in the range $0<k<\mu$. In this range we use identity $n_F(x) =
1-n_F(-x)$. 
The conductivities can then be re-expressed in the form
\begin{align}
 \sigma^B &= \frac{1}{4\pi^2} \int_0^\mu dk\;\,,\label{eq:cmcnice}\\
 \sigma^V = \sigma_\epsilon^B &= \frac{1}{4\pi^2} \int_0^\mu k\; dk\; +
\frac{1}{2\pi^2}
\int_0^\infty x\, n_F\left(\frac x T \right) dx\,,\label{eq:cvcnice}\\
 \sigma^B &= \frac{1}{4\pi^2} \int_0^\mu k^2 dk\; + \frac{\mu}{\pi^2}
\int_0^\infty x\, n_F\left(\frac{x}{T} \right) dx\label{eq:cvcenice}\;\,
\end{align}
These expression have the advantage that the contributions from the thermal
fluctuations are nicely separated from
the finite density contributions.

\subsection{Choice of vacuum}
So far we have described the anomaly related transport for relativistic Weyl
fermions.
The results in \cite{Basar:2013iaa} do however not fit into this scheme
(\ref{eq:cmc})-(\ref{eq:cvce}). If we
assume a vanishing 
CME this would automatically imply a vanishing CVE, whereas \cite{Basar:2013iaa}
find a vanishing
CME but a non-vanishing CVE. Indeed if we assume $\mu_L = \mu_R$ the chiral
magnetic
conductivity  equation (\ref{eq:cmc}) as well as the chiral vortical
conductivity equation (\ref{eq:cvc}) vanish both
(the temperature term cancels upon adding left- and right-handed current).

\begin{figure} [ht]
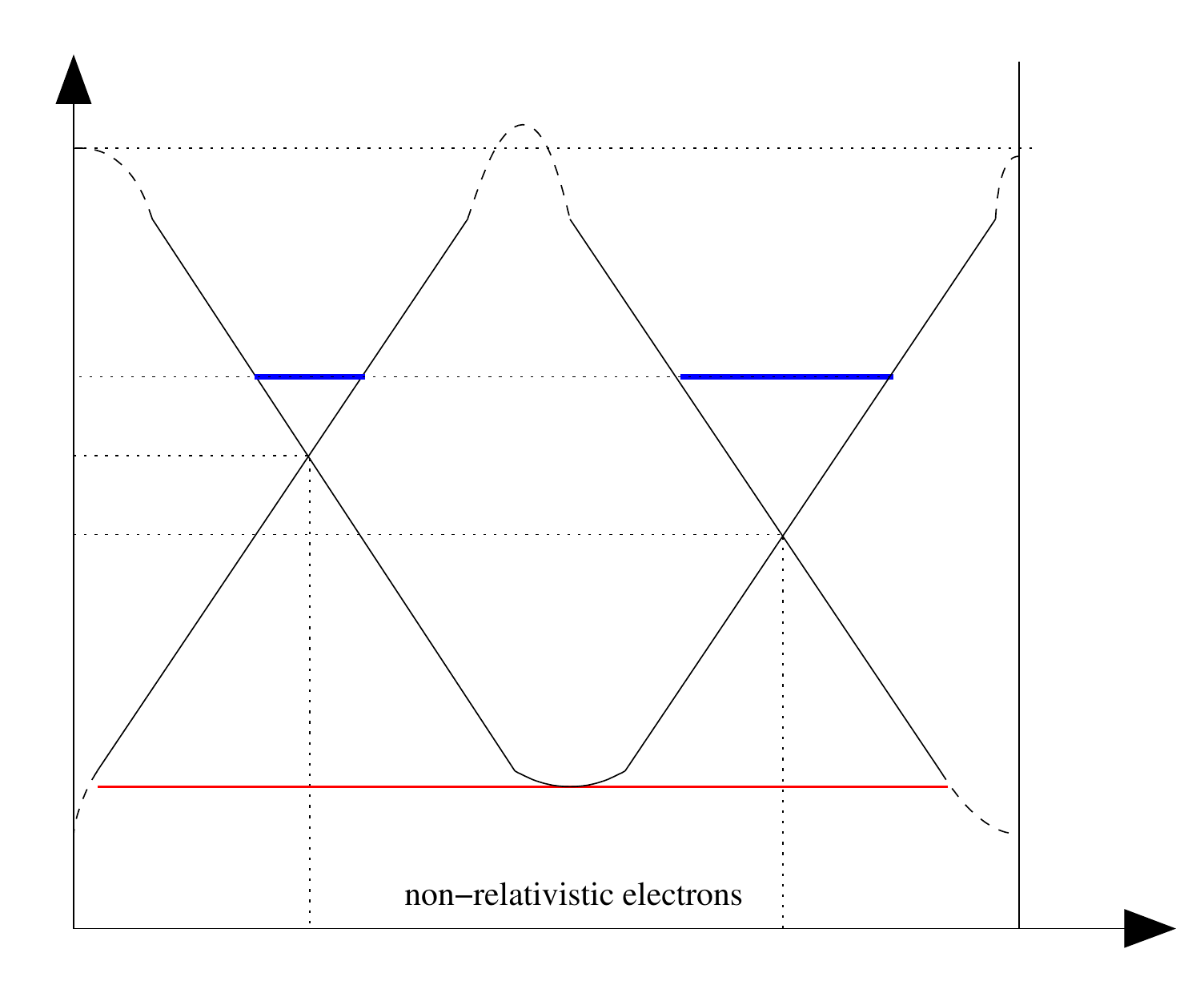
\caption{Schematic depiction of the electronic structure of a Weyl semi-metal.
Two Weyl cones of opposite handedness are located
at $(E_L,k_L)$ and $(E_R,k_R)$. The filling level (chemical potential) is given
by $\mu$ and in equilibrium must be the same
for both cones. Below the level denoted by $E_0$ the description in terms of
Weyl fermions is not valid. 
This provides a natural IR cutoff. From the point of view of the physics of the
excitations near the Fermi surface this is
however better thought of as an UV cutoff since in order to probe it one needs
to create holes (=anti-particles) of energy
$\omega \sim \mu-E_0$. In this sense the low energy effective theory near the
Fermi surface is the one of two Weyl fermions
with chemical potentials $\mu_{R,L} = \mu-E_{R,L}$. The Dirac seas below the
tips of the cones is however not of infinite
depth but reaches down only to $E=E_0$.}
\label{fig:BZone}
\end{figure}

So how can these formulas can be applied to Weyl semi-metals? A schematic
picture of the electronic
structure of a Weyl semi-metal is given in figure \ref{fig:BZone}. 
The energy levels are filled up to the Fermi surface denoted $E=\mu$. The
locations in the Brillouin zone
of the touchings are $(E_R,\vec{k}_R)$ and $(E_L,\vec{k}_L)$. Locally around the
conical band touching points
the Hamiltonian is linear in the momentum $\vec p - \vec{k}_{R,L}$ and takes the
form of right- and left-handed
Weyl equations
\begin{equation}\label{eq:weylH}
 H_{R,L} = E_{R,L} \pm \vec{\sigma}.(\vec p - \vec{k}_{R,L})\,.
\end{equation}
Relative to each other the Weyl cones are displaced in energy by $E_R-E_L$ and
in momentum by $\vec{k}_R- \vec{k}_L$.
An effective action describing this situation is 
\begin{equation}\label{eq:Seff}
 S = \int d^4x\; \bar\Psi\gamma^\mu (i\partial_\mu - \gamma_5 b_\mu)\Psi \,,
\end{equation}
with $E_L-E_R = 2 b_0$ and $\vec{k}_L-\vec{k}_R = 2 \vec{b}$. In the terminology
of the
previous section we can identify the four vector $b_\mu$ with an axial ``gauge''
potential
$A^5_\mu$. But we should not identify $b_0$ with the chemical potential. In
general
the chemical potential is not necessarily a parameter in the Lagrangian, it is
so only
in a specific gauge, e.g. $A_0=\mu$. In the context of the chiral magnetic effect
it has
been shown in \cite{Landsteiner:2012kd} that an analogous gauge choice for the axial
gauge field has to be avoided. We will
therefore define the chemical potential via specific choices of contours in the
complex
frequency plane. However before we do so we need to decide what we consider to
be the quantum field theoretical vacuum.

\begin{figure}[ht]
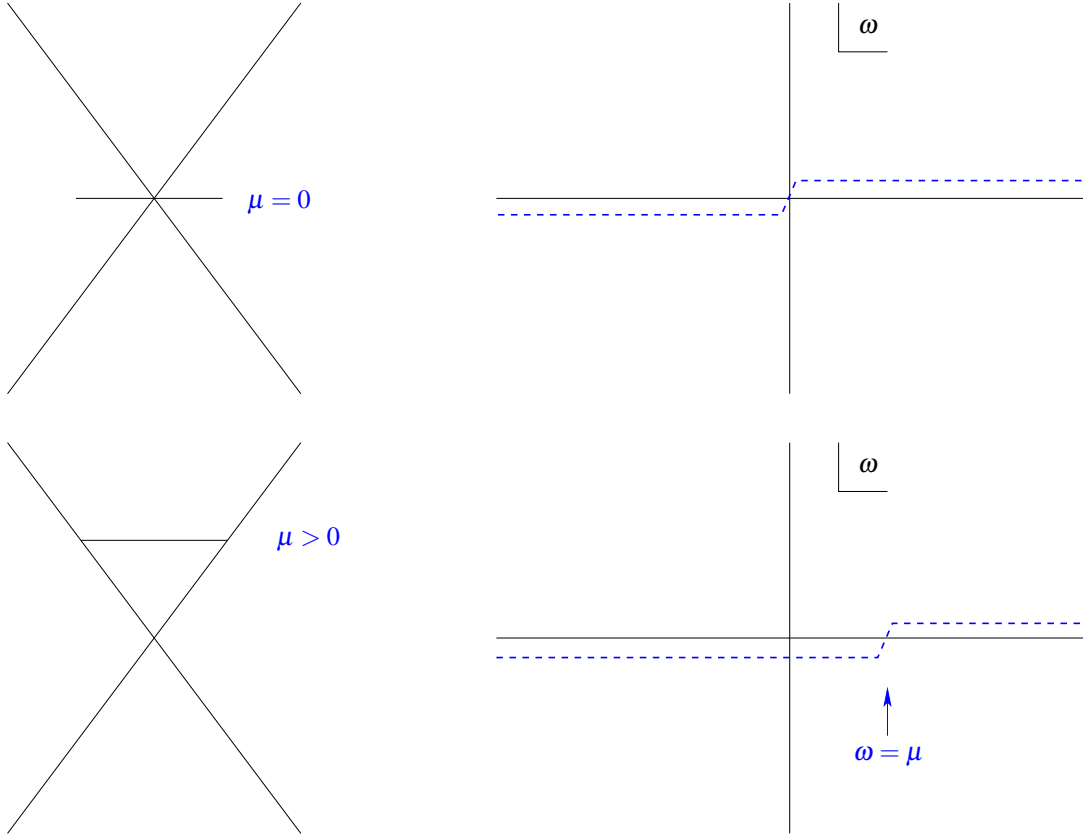
\caption{In the relativistic theory of Weyl fermions the choice of vacuum is
dictated by symmetry and
chosen to lie at the tip of the cone of the dispersion relations. The negative
energy states below are 
reinterpreted as anti-particles via the Feynman boundary conditions. The contour
integration
in the complex $\omega$ plane is shifted slightly below the 
poles corresponding to the negative energy states and slightly above the
positive
energy poles. 
At finite chemical potential
the positive energy states between the vacuum $\omega=0$ and $\omega=\mu$ are
occupied. Because of the
Pauli principle no further states can be created. But a fermion annihilation
operator does create a hole state.
This looks like a negative energy state for the Hamiltonian $H-\mu Q$ which
justifies the gauge choice $A_0=\mu$ normally used in quantum field theory. 
The contour is now shifted below the
poles up to $\omega=\mu$ and 
above the poles for large frequencies. The difference between the two contours
is the contribution of the 
occupied on-shell states at zero temperature.}
\label{fig:vacuum_choice}
\end{figure}

In the relativistic theory of a Dirac fermion we have a unique choice of vacuum
which is
invariant under the discrete symmetries $C,P,T$. It is the usual normal-ordered
vacuum 
which re-interprets negative energy states as propagating backwards in time via
the
Feynman boundary conditions giving rise to negatively charged anti-particles.
This removes 
an a priori infinite Dirac sea of occupied negative energy states. At finite
chemical potential the Feynman
integration contour is shifted by $\mu$. The contribution to an amplitude
stemming from
the occupied on-shell states between $\omega=0$ and $\omega=\mu$ is given by the
difference
of the two contours depicted in figure \ref{fig:vacuum_choice}.

Now from the figure \ref{fig:BZone} it can already be seen that no such
privileged choice of vacuum is available in the case of a Weyl semi-metal. 
In addition it is clear that the validity of the
description of the dynamics of the band electrons
by the Weyl Hamiltonian in equation (\ref{eq:weylH}) is limited to a certain
energy range. Following \cite{Basar:2013iaa} we take $E_0$ to be the lower
cut-off  and $E_1$ the upper cut-off 
at which the description in terms of Weyl fermions breaks down.
Note that from the point of the effective 
theory of Weyl fermions these are both UV cut-offs. This is obvious for $E_1$
but even $E_0$ acts as a UV cutoff since in
order to probe that energy we have to poke a hole in the Fermi surface of the
order of $\mu-E_0$. From the point
of view of the Weyl fermions we have to excite ``anti-particles`` (=hole states)
of energies
$E_{R,L} - E_0$ in the presence of an effective chemical
potential $\mu_{R,L} = \mu - E_{R,L}$. We will for simplicity work with sharp
cutoffs $E_{0,1}$. This is most likely a drastic oversimplification. In a real
Weyl semi-metal the change of the dispersion relation from relativistic to
non-relativistic
will by gradual. We think that this situation can however be modeled by
effective
(model dependent) cut-offs $E_{0,1}$. A rough estimate is given by the energy
level at
which the dispersion obtains a local minimum as in figure
(\ref{fig:BZone}).\footnote{A more realistic model would be to include a (small)
mass term in (\ref{eq:Seff}) and work out the values of the anomalous transport
coefficients
via Kubo formulae \cite{Landsteiner:2012kd}.}.
Below $E_0$ we do not even have the relativistic
vacuum of the Weyl fermions present. The formulas (\ref{eq:cmc}), (\ref{eq:cvc})
and (\ref{eq:cvce}) however are
derived assuming the description in terms of Weyl formulas to be valid down to
arbitrarily low energies and by
subtracting a suitable chosen vacuum (an infinitely deep Dirac sea). In Weyl
semi-metals the Dirac sea is however physical and rather 
rather shallow, it reaches down from $E_{L,R}$ to $E_0$ and its contribution has
to be taken into account properly. Because of this we suggest to
define the ''vacuum`` at the energy $E_0$. In that way we include all occupied
electron states in the Weyl cones even if they are below the tip of the cone.

\begin{figure} [ht]
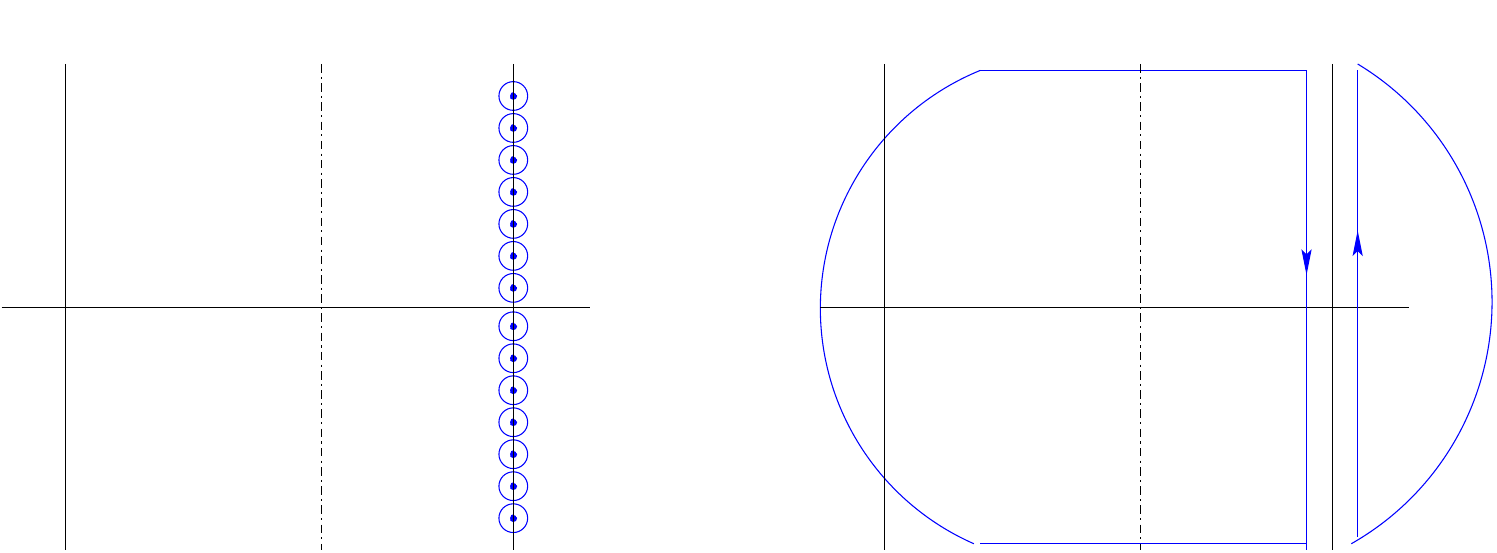
\caption{Contour integration for the summation over Matsubara frequencies for a
generic one-loop amplitude at
finite chemical potential $\mu$. The finite temperature contribution is captured
by the sum over the residues 
weighted by the Fermi-Dirac distribution inside the two $D$ shaped contours in
the right figure.}
\label{fig:contour2}
\end{figure}

\begin{figure} [ht]
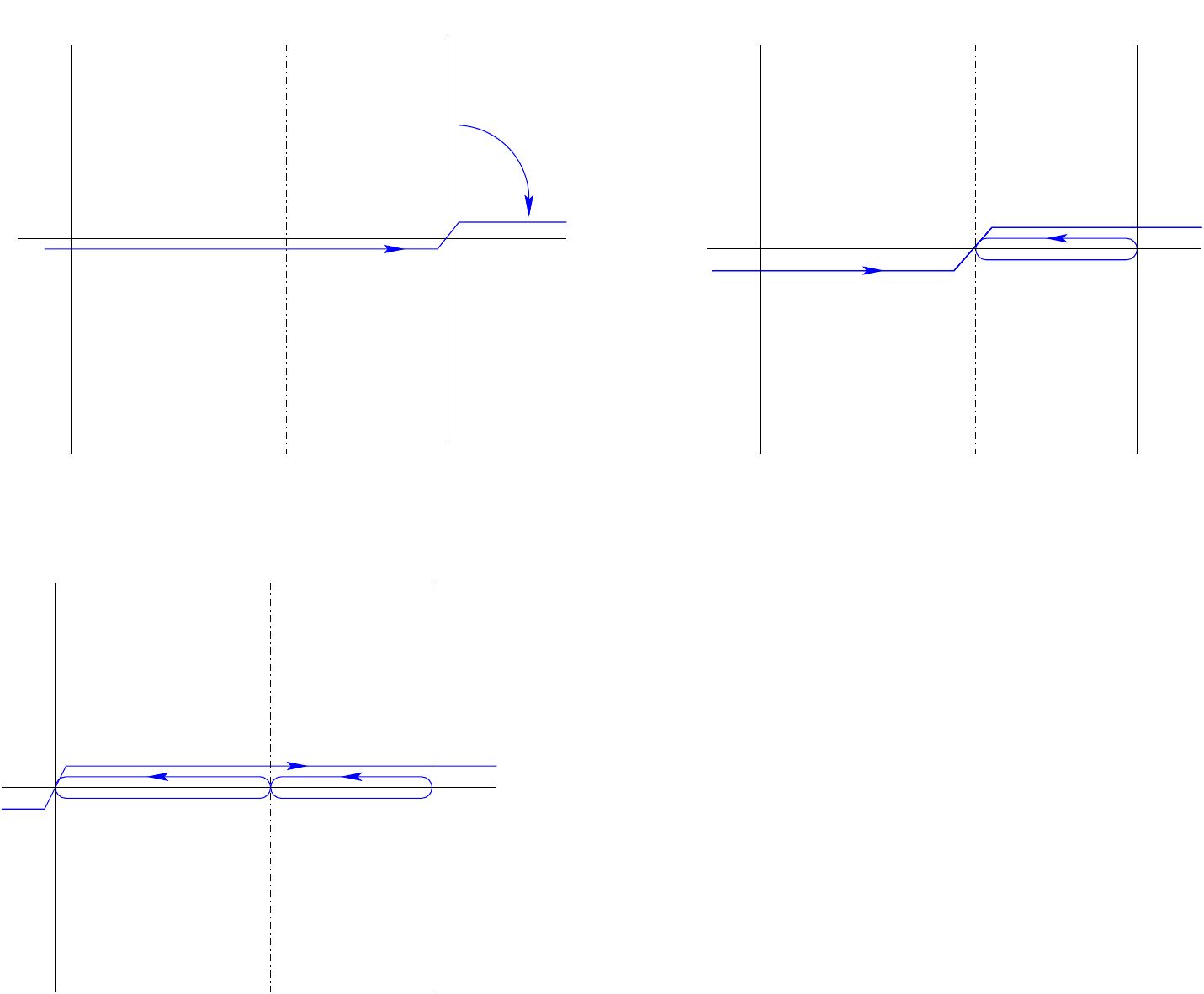
\caption{Wick rotation and contour deformation for the non-compact vacuum
contour. The contour at $E=\mu$ 
describes particles for $z>\mu$ and the holes as anti-particles for $z<\mu$. In
the second figure
we have the vacuum contour of the usual relativistic Weyl fermion with
anti-particles below $z<\mu-E_{R,L}$ and
particles above and the occupied particle states between $E=E_{R,L}$ and
$E=\mu$. In the third part of the 
figure we have the vacuum down at $E=E_0$ and the states occupied between
$E=E_0$ and $E=E_{R,L}$.}
\label{fig:contour3}
\end{figure}

This is implemented by the following contour deformation for a
generic one-loop amplitude
at finite chemical potential and finite temperature.
Using standard manipulations we can write the sum over Matsubara frequencies for
a generic function $f(z)$ as
\begin{align}
 T\sum_{n} f(i\omega_n+\mu) &= \frac 1 2 \int_{\mathcal{C}} f(z) \tanh\left(
\frac{z-\mu}{2T} \right) \frac{dz}{2\pi i}\,,\\ 
&= \int_{\mu-i \infty}^{\mu+i \infty} f(z)  \frac{dz}{2\pi i} +
\sum_{\Re(z_i)>\mu} res(f)(z_i) n_F\left( \frac{z_i-\mu}{T}\right) -  
\sum_{\Re(z_i)<\mu} res(f)(z_i) n_F\left( \frac{\mu-z_i}{T}\right) \,,
\end{align}
where the contours are as in figure \ref{fig:contour2} and $z_i$ are the poles of $f(z)$.
We have used $\tanh(x) = 1- 2 n_F(2x)$ for $\Re(z)>\mu$ and   $\tanh(x) = - 1+ 2
n_F(-2x)$ for $\Re(z)<\mu$.
This allows to extract
the finite temperature contribution that comes from the fluctuations around the
Fermi surface at $ E=\mu$ for a general amplitude just as what we did
explicitly for the conductivities in  (\ref{eq:cmcnice}-\ref{eq:cvcenice}).
The thermal contribution vanishes at $T=0$. Furthermore for $\Re(z_i)\gg \mu$ 
the thermal contributions are exponentially suppressed.
The temperature independent integral can be Wick rotated and deformed into the
``vacuum'' contour at $E=E_0$ (see
figure \ref{fig:contour3}). We will ignore this vacuum contour for the moment
but
comment on its significance in the next section.

The remaining contour
integration is over a compact cycle and can be decomposed as the sum of
$\mathcal{C}_{\mu-E_i}$ and $-\mathcal{C}_{-E_i}$. 
The latter contours are the same as for a Weyl fermion at chemical potential
$\mu_{eff} = \mu - E_i$ and the contour
of another Weyl fermion at effective chemical potential $\mu_{eff} = E_0-E_i$
were it not for the orientation of
the contour. Because of this orientation we need to subtract this contribution
rather than to add it.

These considerations lead to the general formula for the anomalous transport in
Weyl semi-metals
\begin{equation}\label{eq:wsmat}
 \sigma_{wsm} = \sigma(\mu-E_i,T) - \sigma(E_0-E_i, 0)\,.
\end{equation}

\subsection{Anomalous Conductivities}\label{sec:anomalconds}
Let us now apply this formula to the Chiral Magnetic Effect. For each isolated
Weyl cone we find
\begin{equation}\label{eq:cmeLR}
 \vec{J}_{R,L} = \pm\left(\frac{\mu-E_{R,L}}{4\pi^2} -
\frac{E_0-E_{R,L}}{4\pi^2}\right)\vec{B}  = \pm\frac{\mu-E_0}{4\pi^2} \vec{B}\,.
\end{equation}

Therefore the (covariant vector-) current $J_R + J_L$ vanishes and there is no
CME in a
Weyl semi-metal in equilibrium!  While we find that there is not chiral magnetic
current
we will see however in the next section that the anomaly will still give rise to
a charge
separation effect once the effects from the edges are taken into account.
On the other hand the Chiral Separation Effect(CSE), i.e. the generation of an
axial current \footnote{In the condensed matter
literature this is usually referred to as ``valley`` current.}, is fully
realized 
\begin{align}
 \vec{J} &= 0\,,\\
 \vec{J}_5 &= \frac{\mu-E_0}{2\pi^2} \vec{B}\,.
\end{align}
The driving force for the CME is an imbalance in the
Fermi surfaces of left- and right-handed fermions \cite{Kharzeev:2007jp}.
It has been suggested that this can be achieved by placing the system in
parallel electric and magnetic fields \cite{Son:2012wh, Kharzeev:2012dc} which
will
pump such an imbalance into the Weyl semi-metal via the anomaly equation
(\ref{eq:anomaly}). We will discuss another way of at least locally achieving an
such an imbalance in section \ref{sec:discussion}. 

Now we turn to the chiral vortical effect. According to our formula
(\ref{eq:wsmat}) it is given by
\begin{align}\label{eq:cvcwsm}
 \vec{J}_{R,L} &= \pm \left( \frac{(\mu-E_{R,L})^2 }{4\pi^2} +
\frac{T^2}{12}- \frac{(E_0-E_{R,L})^2}{4\pi^2} \right) \vec{\omega} \,,\\
&= \pm \left( \frac{(\mu-E_0)(\mu+E_0-2E_{R,L}) }{4\pi^2} + \frac{T^2}{12}
\right) \vec{\omega}
\end{align}
The temperature independent part coincides exactly with the result of
\cite{Basar:2013iaa}.
The vortical effect in the electric current is  given by
\begin{equation}
 \vec{J} = \frac{(E_L-E_R)(\mu-E_0)}{2\pi^2} \vec\omega\,.
\end{equation}

Let us now turn to the energy current. The general formulas (\ref{eq:cmc}) -
(\ref{eq:cvce}) suggest also anomalous transport
in the energy current. 
In particular our model predicts an energy current in a magnetic field of
\begin{equation}
  \vec{J}_{\epsilon,R,L} = \pm \left( \frac{(\mu-E_0)(\mu+E_0-2E_{R,L})
}{8\pi^2} + \frac{T^2}{24} \right) \vec{B}\,.
\end{equation}
The equality of the Chiral Magnetic Conductivity for the energy current and the
Chiral Vortical Conductivity
for the charge current follows from the Kubo formulae \cite{Landsteiner:2012kd}.

However, this is not yet the total energy flow generated. The tips of the cones
are located at $E=E_{R,L}$ and
they can be understood as background values for left-handed and right-handed
temporal components of gauge fields.
They contribute terms to the action of the form
\begin{equation}
 \int A_\mu^{R,L} J^\mu_{R,L} \,.
\end{equation}
The energy-momentum tensor can be understood as the reaction of the system to a
variation in the metric. In particular
if we switch on a mixed time-spatial component $h_{0i}$ in the metric we induce
a term\footnote{There is also the term $k_{R,L}^i J_{R,L}^0 h_{0i}$ which
might give additional contributions if $J^0$ is not vanishing.}
\begin{equation}\label{eq:drift}
 \int A^0 j^i h_{0i} = \int E_{R,L} j^i_{R,L} h_{0i}\,.
\end{equation}
and therefore there is the additional ''drift`` in the energy flow of $E_{R,L}
\vec{j}_{R,L}$.
For each Weyl cone we find thus
\begin{align}
\vec{J}_{\epsilon,R,L} &= \pm \left( \frac{(\mu-E_{R,L})(\mu+E_0-2E_{R,L})
}{8\pi^2} + \frac{T^2}{24} + \frac{E_{R,L}(\mu-E_0)}{4\pi^2}\right)\vec{B}\,.\\
 & = \pm\left(\frac{\mu^2-E_0^2}{8\pi^2} + \frac{T^2}{24} \right)\vec{B}\,.
\end{align}
The sum of the contributions of the two Weyl cones vanishes.

Let us compare this now with kinetic theory. The energy flow in one Weyl cone we
compute as
\begin{equation}
 \vec{J}_{\epsilon} = \int \frac{d^3p}{(2\pi)^3} \sqrt{G} n_F(E) \,E \vec{\dot{x}} \,,
\end{equation}
where $G = 1 + \vec \Omega . \vec B$ is the phase space measure in presence
of the Berry curvature and magnetic field \cite{Son:2012wh,Stephanov:2012ki}.
And using the steps outlined in \cite{Basar:2013iaa} this evaluates to
\begin{align}
 \vec{J}_{\epsilon} &= \pm\frac{1}{4\pi^2} \vec{B} \int_{E_0}^\infty E n_F(E) dE
\,,\\
&= \pm \frac{\mu^2 - E_0^2}{4\pi^2}\vec{B} + O(T)\,.
\end{align}
Where in the last line we have taken the zero temperature limit. Again this
agrees with our Weyl fermion model.

Finally let us also compute the energy flow due to rotation.  According to what
we outlined before the
response in $T^{0i}$ is the one that does not take into account the energy drift
due to equation (\ref{eq:drift}).
Our Weyl fermion model predicts (ignoring for a moment the drift term)
\begin{align}
 \vec{J}_{\epsilon,R,L} &= \pm \left( \frac{(\mu-E_{R,L})^3}{6\pi^2} +
\frac{(\mu-E_{R,L}) T^2}{6} - \frac{(E_{R,L}-E_0)^3}{6\pi^2} \right)
\vec{\omega}\,.
\end{align}
We can also subtract the drift term in the kinetic model. It corresponds to
inserting $E-E_{R,L}$ instead of $E$ in the phase space integrals. We use also
the 
heuristic substitution $\vec B = 2 (E-E_{R,L}) \vec{\omega}$
suggested in \cite{Basar:2013iaa,Stephanov:2012ki} to find the kinetic
theory based expression
\begin{align}
  \vec{J}_{\epsilon,R,L} &= \pm \vec{\omega} \frac{1}{2\pi^2} \int_{E_0}^\infty
dE (E-E_{R,L})^2 n_F(E) \,,\\
&=  \pm \left( \frac{(\mu-E_{R,L})^3}{6\pi^2} - \frac{(E_0-E_{R,L})^3}{6\pi^2} +
O(T) \right) \vec{\omega}\,.
\end{align}
Again we find agreement with the Weyl fermion model. 
For completeness we also state the result for the total energy flow due to
rotation including the drift terms
\begin{align}
 \vec{J}_{\epsilon} = \left(\frac{(\mu^2-E_0^2) (E_L-E_R)}{4\pi^2 } +
\frac{(E_L-E_R) T^2}{12} \right)\vec{\omega}
\end{align}
This result is of some significance since it is the only vector like (i.e. sum
as opposed to difference of right-
and left-handed contributions) current sensitive to the $T^2$ temperature
dependence and therefore to the
mixed gauge-gravitational anomaly.

\section{Anomaly and CME at the edge}
\label{sec:covcons}

In this section we will try to connect the formula (\ref{eq:anomaly}) with
standard field theoretical language.

In the quantum field theory of Weyl fermions there are at least two conventions
on how to phrase the anomaly. The first one, called the {\em consistent}
anomaly, defines
the current as the functional derivative of the effective action w.r.t. to the
gauge field.
The anomaly comes from a triangle diagram with three currents at the vertices.
For
the moment we think of having only one (right-handed) Weyl fermion, all three currents are the
same and
that imposes a Bose symmetry of the vertices. In this case the anomaly is given
by \cite{Bertlmann:1996xk, Fujikawa:2004cx}
\begin{equation} \label{eq:anomalycons}
\partial_\mu J_{\mathrm{cons}}^\mu =  \frac{1}{96\pi^2}
\epsilon^{\mu\nu\rho\lambda} F_{\mu\nu}F_{\rho\lambda} = \frac{1}{12\pi^2} \vec E \cdot \vec B\,.
\end{equation}
If we compare this to (\ref{eq:anomaly}) we see that this gives a anomaly
coefficient
that is smaller by a factor of $1/3$. The deeper reason is that because of the
Bose symmetry
at the vertices the anomaly is distributed equally amongst them.

There is however another definition of the current that does not follow from
functional differentiation
of the effective action. Instead we could define the current as a
gauge-invariant operator, e.g. by 
gauge-covariant point splitting. The covariant current is gauge invariant even if the
gauge transformations are effected by an anomaly.
The quantum operator $J^\mu_{\mathrm{cov}}$
defined in this way obeys the {\em covariant}
anomaly equation \cite{Bertlmann:1996xk, Fujikawa:2004cx, Bardeen:1984pm}
\begin{equation} \label{eq:anomalycov}
\partial_\mu J_{\mathrm{cov}}^\mu =  \frac{1}{32\pi^2}
\epsilon^{\mu\nu\rho\lambda}F_{\mu\nu}F_{\rho\lambda} = \frac{1}{4\pi^2} \vec E\cdot \vec B\,.
\end{equation}
The two definitions of currents differ by a Chern-Simons current
\begin{equation}
 J^\mu_{\mathrm{cov}} = J^\mu_{\mathrm{cons}} + \frac{1}{24\pi^2}
\epsilon^{\mu\nu\rho\lambda} A_\nu F_{\rho\lambda}\,.
\end{equation}

Now the important point is that the kinetic definition of the current obeys the
covariant anomaly equation. Therefore we should
identify the kinetic current with the covariant current of quantum field
theory. 

Let us now add a left-handed and a right-handed fermion. The covariant vector
and axial currents are simply defined
as sum and difference of right-handed and left-handed currents. We also
introduce vector-like gauge fields $A_\mu$ 
that couple with the same sign to left- and right-handed fermions and axial
gauge fields that couple with opposite
signs to left- and right-handed fermions. The anomalies in these {\em covariant}
currents can be written as
\begin{align}
 \partial_\mu J^\mu_{V,\mathrm{cov}} &= 
\frac{1}{2\pi^2} \left(\vec E_5 \cdot \vec B + \vec E \cdot\vec B_5\right)\,\\
 \partial_\mu J^\mu_{5,\mathrm{cov}}& 
= \frac{1}{2\pi^2} \left(\vec E \cdot\vec B + \vec E_5
\cdot\vec B_5\right)\,
\end{align}

The important lesson is that the vector current, i.e. the sum of left-handed and
right-handed fermion number currents, is {\em not}
conserved! 

Let us see now what consequence this non-conservation has if we have a finite
region in space in which the temporal component
of the axial gauge field is switched on. In the case of the Weyl semi-metal we
see from the effective action (\ref{eq:Seff}) that
the energy difference of left- and right-handed cones $1/2(E_R-E_L)$ plays the
role of $A^5_0$! We will model therefore
the Weyl semi-metal as a region of space in which $A^5_0\neq 0$ and constant. So
for a cubic body of extensions $\{L_1, L_2, L_3\}$ 
\begin{equation}
 A^5_0 = \frac{E_R-E_L}{2}\prod_{i=1}^3 \left[ \Theta\left(
x_i+\frac{L_i}{2}\right) - \Theta\left( x_i-\frac{L_i}{2}\right)\right]\,.
\end{equation}
If we apply a magnetic field in the $3$-direction it follows from
(\ref{eq:anomalycov})  and the fact that there is no
bulk current that at the edges 
\begin{equation}
 \partial_t \rho |_{x_3=\pm L_3/2} = \pm \frac{E_R-E_L}{4\pi^2} B \,.
\end{equation}
Although in this description based on the covariant current we
found that there is no bulk chiral magnetic current, there is effective
accumulation of electric charges on the edges of the 
Weyl semi-metal because of the anomaly in the covariant current! We note that no net
charge is created and this is guaranteed as long as
at spatial infinity we are dealing with the trivial vacuum, i.e. the one in
which the axial gauge fields are zero! Indeed, since the anomaly is a total derivative
\begin{equation}
 \int_{\Omega} \partial_t \rho = \frac{1}{4\pi^2} \int_\Omega \epsilon^{ijk} \left( \partial_i A^5_0 F_{jk} \right) =
\frac{1}{4\pi^2} \int_{\partial\Omega} dS_i \epsilon^{ijk}  \left( A^5_0 F_{jk} \right)\,,
\end{equation}
and we can take the surface bounding the volume $\Omega$ to lie outside the region where $A^5_0$ is non-zero.

We also note that this is precisely the charge that would accumulate at the
edges if there was a bulk current 
\begin{equation}\label{eq:notcme}
 \vec{J} = \frac{E_L-E_R}{4\pi^2} \vec B 
\end{equation}
This looks like the chiral magnetic effect but it its origin is rather
different. It is present even if there is no genuine axial
chemical potential, i.e. if there is no imbalance in the number of right-handed
and left-handed occupied fermion states. 
Chiral kinetic theory which is concerned only with the occupied on-shell states
within the local Weyl cones does not have this
type of bulk current. However the kinetic (covariant) current is anomalous and
therefore it is the anomaly that induces the edge
charges. We interpret this phenomon as follows: due to the anomaly the Fermi
level gets diminished at one edge whereas it
expands at the other edge. In the bulk in between that necessitates a
reshuffling of the electronic states below our cutoff 
$E_0$ and this is the origin of the current (\ref{eq:notcme}). 
In our simple model with hard cut-off at $E=E_0$ we have no direct access to 
this reshuffling of states.  
In quantum field theory with a hard cutoff this current appears if one imposes a
 regularization scheme that preserves
invariance under the gauge transformations associated to the gauge field $A_\mu$
as has
first been noted in \cite{Gynther:2010ed}. In fact if one does not calculate the
covariant current but rather the consistent
current in a scheme that preserves vector-like gauge invariance one finds the
{\em consistent} current \cite{Jackiw, Bardeen:1969}
\begin{equation}\label{eq:conscurrent}
 J^\mu_{V,\mathrm{cons}} = J^\mu_{V,\mathrm{cov}} - \frac{1}{4\pi^2}
\epsilon^{\mu\nu\rho\lambda} A^5_\nu F_{\rho\lambda}\,.
\end{equation}
As we have argued here it is however not necessary to introduce the consistent
and exactly conserved current (\ref{eq:conscurrent}).
We can work with the covariant current more natural to kinetic theory but have
to take into account the anomaly at the edges\footnote{The
covariant current can of course not be coupled directly to a gauge field obeying
Maxwell's equations since $J^\mu = \partial_\nu F^{\mu\nu}$
entails that the current must be conserved. Therefore it is better to think of
the covariant current to be the source of a Chern-Simons
modification of Maxwell's equations $J^\mu_{\mathrm{cov}} = \partial_\mu
F^{\mu\nu} + \frac{1}{4\pi^2}\epsilon^{\mu\nu\rho\lambda} A^5_\nu
F_{\rho\lambda}$.} .

Since vorticity does not source the anomaly (\ref{eq:anomalycov}) (a constant
vorticity does not excite the higher derivative gravitational anomaly) there is no
such edge contribution present in the chiral vortical effect!

\section{Experimental test of CME and CVE}
\label{sec:experiments}
Now we are going to suggest an experiment that can test some of the main
points
discussed in the previous sections. We ask how can we induce the chiral
magnetic
effect and at the same time test for the $T^2$ term related to the gravitational
contribution
to the axial anomaly. 

\begin{figure}
 \includegraphics[scale=0.35]{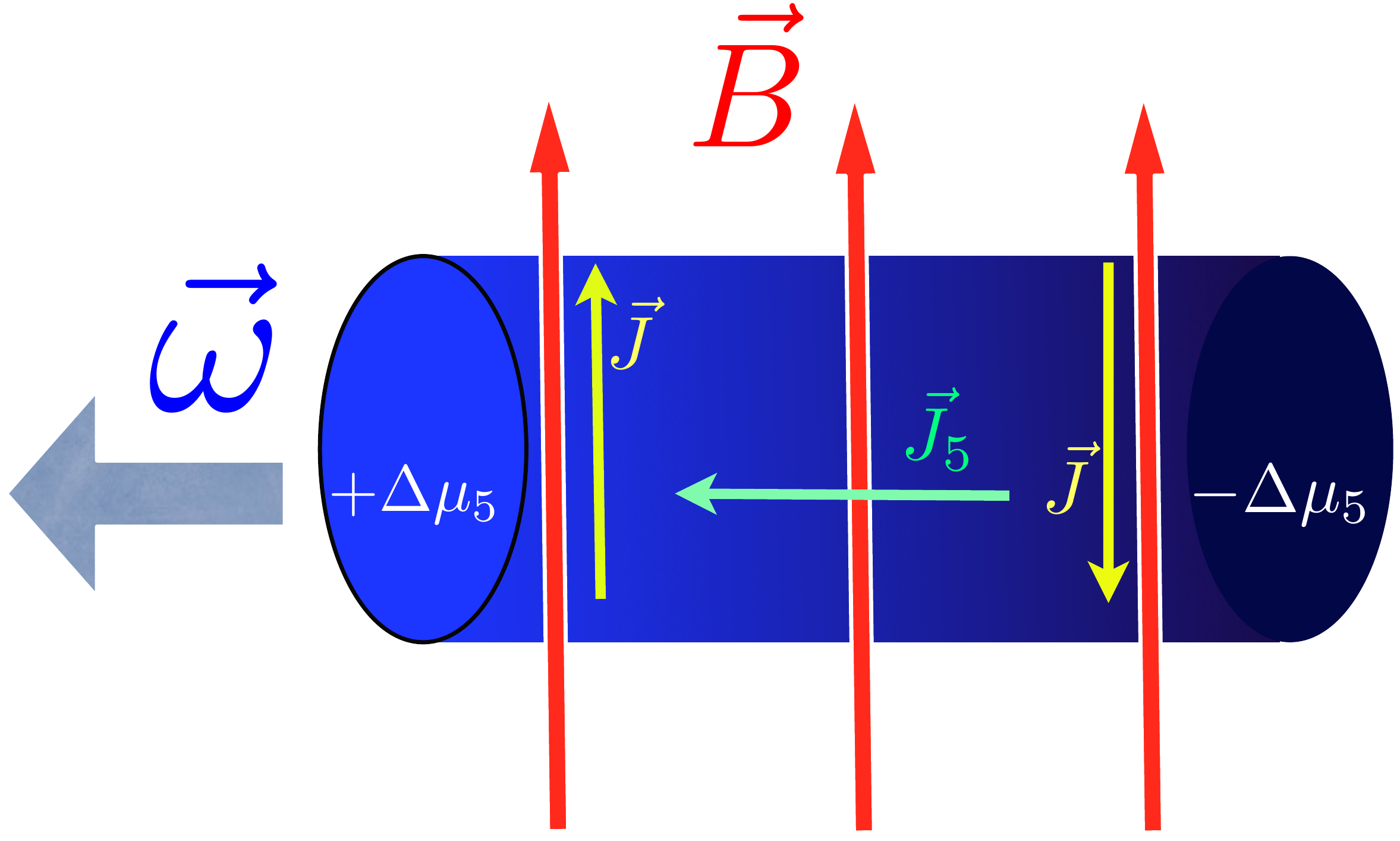}
\caption{It should be possible to test the CVE and CME at once with a cylinder
shaped Weyl semi-metal.
The rotation will induce an axial current that will lead to a stationary state
with effective axial
chemical potentials of equal magnitude but opposite sign at the ends of the
cylinder. A magnetic
field will induce an electric current that depending on the sign of the local
axial chemical potential will
either flow along the magnetic field or contrary to it.\label{fig:cvecme}}
\end{figure}

\begin{figure}
 \includegraphics[scale=0.45]{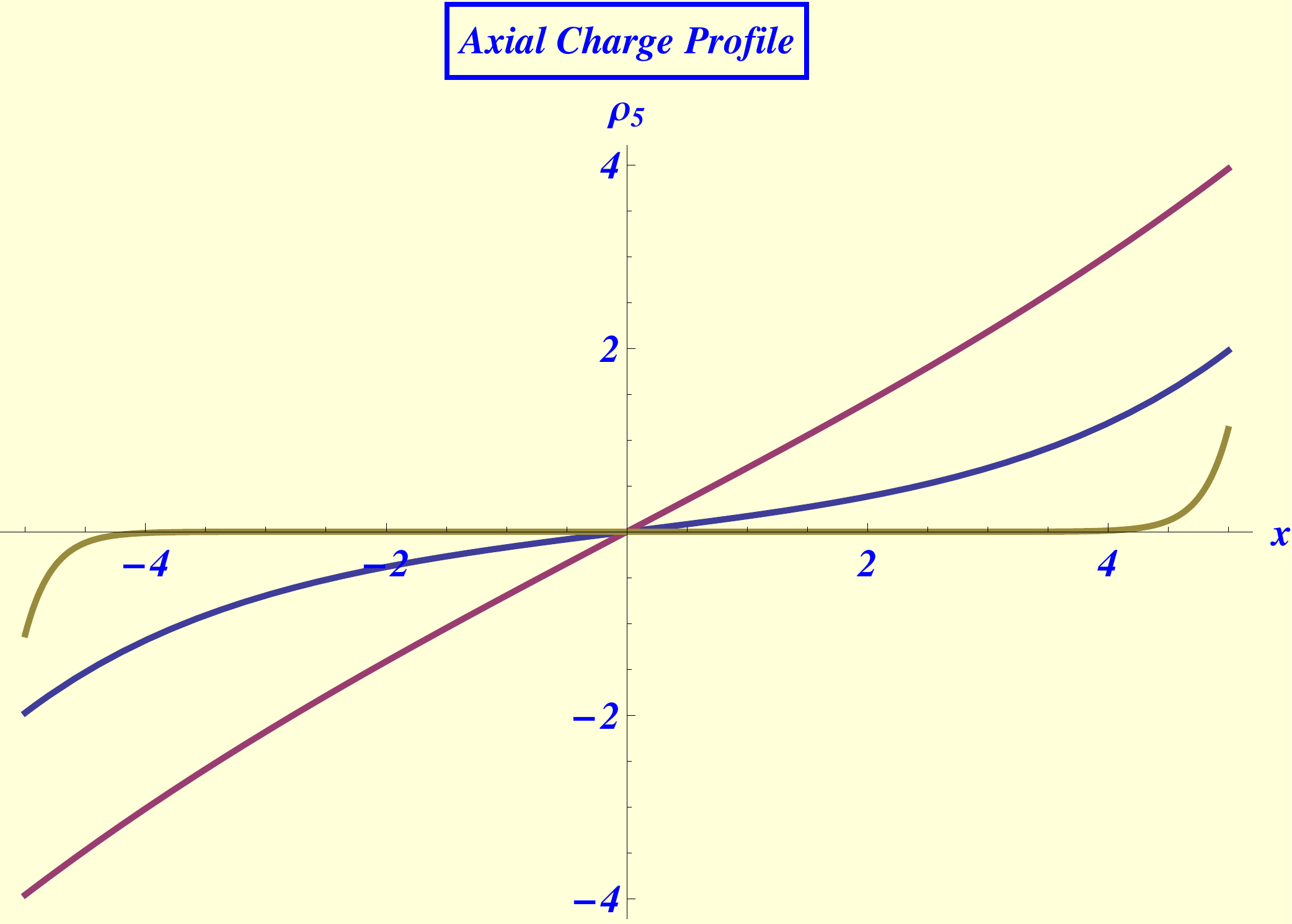}
\caption{The figure shows the axial charge profile along the rotating cylinder
induced via the chiral vortical effect in
the axial (valley) current. For long lived chiral quasiparticles diffusion
dominates and the profile is simply a straight line.
For short lifetimes or long cylinders decay in the bulk dominates and we get
axial charge accumulation only at the edges. 
The blue curve shows and intermediate case. [Color online] 
\label{fig:chargeprofile}}
\end{figure}

According to (\ref{eq:cvcwsm}) rotation gives rise to an axial current. 
That is true even when $E_R=E_L$. 
This fact could be used to perform an experiment that at once tests both
the CME and the CVE. If we have a cylinder of Weyl semi-metal and we rotate it
along its axes the vortical CSE will lead to the production of local imbalances 
between the right-handed and left-handed Fermi surfaces.
More precisely, the rotation will accumulate axial charge on one side of the
cylinder and deplete it on the other side. 
We want to take into account that in a real situation axial charge will not
be conserved. We take $\tau$ to be the lifetime of a chiral quasiparticle.
The continuity equation for axial charge is then modified to 
\begin{equation}\label{eq:axialnonconservation}
 \partial_t \rho_5 = -\frac 1 \tau \rho_5 - \vec\nabla\vec J_5\,.
\end{equation}
Combining the diffusion law with the chiral vortical conductivity we
find the axial current to be
\begin{equation}\label{eq:axialdiffusion}
 \vec{J}_5 = - D_5 \vec\nabla \rho_5 + \sigma_V \vec{\omega}\,,
\end{equation}
where $D_5$ is the diffusion constant for axial charge. We assume now that
$\rho_5$ depends only on the coordinate along the cylinder which we denote by $x$.
Since the divergence of the vorticity vanishes we can combine these two
equations to 
\begin{equation}
 \partial_t \rho_5 = -\frac 1 \tau \rho_5 + D_5 \partial_x^2 \rho_5\,.
\end{equation}
To find a stationary state we look  for time independent solutions obeying
the boundary conditions that the current (\ref{eq:axialdiffusion}) vanishes
at the edges of the cylinder. We take the longitudinal extension to be $L$.
The solution is then given by
\begin{equation}\label{eq:axialprofile}
\rho_5(x) =\sigma_V \omega\frac{\sqrt{\tau }}{\sqrt{D_5}}  
\sinh \left(\frac{x}{\sqrt{D_5 \tau }}\right)\text{sech} \left(\frac{L}{2\sqrt{D_5 \tau }}\right)
\end{equation}
in coordinates in which the cylinder ends are at $x=\pm\frac L 2$.
Applying now a perpendicular magnetic field produces an electric current via the
chiral magnetic
effect in the bulk (as opposed to the edge effect described in section
\ref{sec:covcons}.
The profile of the electric current along the cylinder should be directly
proportional to the profile of the axial charge
since this current is 
\begin{equation}J_z(x) = \frac{ \rho_5(x)}{ (\chi_5 2\pi^2)} B_z\,,
 \end{equation}
with $\chi_5$ the axial charge susceptibility.
In figure \ref{fig:chargeprofile} we plot the axial charge profile for three
cases, the diffusion dominated case
in which the lifetime of the chiral quasiparticles is long, an intermediate
regime and a regime that is dominated
by axial charge decay. In the latter the axial charge will be confined to a
narrow region at the edges of order
$\delta x \sim \sqrt{D_5 \tau}$. Note that the decay will always dominate for
long cylinders.

 This effect depends
of course on the material constants $\tau$, $D_5$ and $\chi_5$ whose
determination goes beyond the purpose of this paper. Regardless of the 
their precise values  the effect is directly proportional to the chiral
vortical conductivity in the axial (valley) current.
We find this effect particularly interesting because this conductivity does
depend on temperature $\sigma_V = const. + \frac{T^2}{6}$.
As outlined previously the temperature dependence of the anomalous transport
phenomena can be understood to be a direct consequence
of the gravitational contribution to the axial anomaly. From the point of view of
high-energy particle physics the gravitational anomaly
describes the decay of a neutral pion into two gravitons, which due to the
weakness of the gravitational interactions at the energy scales
available at colliders such as LHC is impossible to observe. Therefore the
anomalous transport properties of Weyl semi-metals offer a unique
and exciting opportunity to connect such a subtle quantum effect as the
gravitational anomaly to phenomena observable in the laboratory.

\section{Discussion}
\label{sec:discussion}

We have presented a simple field theoretical model for anomaly related transport
in Weyl semi-metals.
The basic idea was to model the Weyl semi-metal as far as possible by the known
physics of Weyl fermions. Deviations arise because the choice of vacuum is
not dictated by symmetries. This forces us to consider the contributions of
the finite Dirac sea, following in this point \cite{Basar:2013iaa}.
The resulting conductivities for the covariant currents follow the law
(\ref{eq:wsmat}). 
Where they overlap our results agree with the recent results derived using
kinetic 
theory \cite{Basar:2013iaa}. In particular the chiral magnetic current
vanishes whereas the chiral vortical current does not.  The energy current follow
this
pattern.
The chiral separation effect on the other hand is fully present. 

The temperature dependencies are given by a simple $T^2$ scaling law.
This
assumes of course that
the thermal fluctuations follow the spectrum present inside the Weyl cones. If
the temperature gets
too high, i.e. either of the order of $T\approx E_1-\mu$ or $T\approx \mu-E_0$,
i.e. if the
temperature is high enough to probe the cutoffs, we expect deviations from the
simple $T^2$ behavior.

Even though we found a vanishing chiral magnetic current in the bulk of the 
Weyl semi-metal we argued that there is still an anomaly induced charge accumulation
at the edges, positive on one side and negative on the other. 
This arises because we based our discussion on the {\em covariant} current. We also
argued that it is this quantum definition of current that should be identified with
the current of kinetic theory. We also showed that the total charge is conserved
and that the local charge production rate can be re-written as and inflow of charge
via a Chern-Simons current not contained in the kinetic description. 

We have also discussed how the combination of the CME and the CVE in the axial current
can be used to test for the presence of the $T^2$ term related to the gravitational anomaly.
This constitutes an independent experimental setup to the already proposed  in
\cite{Chernodub:2013kya}.

Finally we note that in a complete treatment  also the anomalous Hall effect
should be included and
the distinction between covariant and consistent current might also be of
relevance there. We leave
this problem for future investigation.

\bigskip
\acknowledgments{
I have profited from discussions with the participants
of the ECT* workshop
``QCD in strong magnetic fields'', Nov. 12-16 November 2012, Trento, Italy and
and the INT program ``Gauge field dynamics in and out of equilibrium'',
University of
Washington, Seattle, USA.
I would like to thank G. Basar, M. Chernodub, A. Cortijo, D. Kharzeev, A.
Grushin, H.-U. Yee, M. Vozmediano
for most enjoyable and helpful discussions.
This work has been supported by MEC and FEDER grants FPA2009-07908 and
FPA2012-32828,
Consolider Ingenio Programme CPAN (CSD2007-00042), Comunidad de Madrid HEP-HACOS
S2009/ESP-1473
and MINECO Centro de excelencia Severo Ochoa Program under grant SEV-2012-0249.
}

\end{document}